\documentclass{article}
\usepackage{setspace}
\usepackage{amsmath, amsthm, amsfonts,setspace,graphicx}
\usepackage[usenames, dvipsnames]{color}
\usepackage{fullpage}
\usepackage{dsfont}
\usepackage{natbib}
\usepackage{booktabs} 
\newcommand\independent{\protect\mathpalette{\protect\independenT}{\perp}} 
\def\independenT#1#2{\mathrel{\rlap{$#1#2$}\mkern2mu{#1#2}}} 

\newcommand{\E}{\mathbb{E}}

\begin{document}

\title{Causal inference from observational data: \\ Estimating the effect of contributions on visitation frequency at LinkedIn}

\author{Iavor Bojinov, Ye Tu, Min Liu, and Ya Xu \\
\textit{LinkedIn}\\
Contact email: \texttt{iav.bojinov@gmail.com}}

\maketitle






\begin{abstract}
Randomized experiments (A/B testings) have become the standard way for web-facing companies to guide innovation, evaluate new products, and prioritize ideas. There are times, however, when running an experiment is too complicated (e.g., we have not built the infrastructure), costly (e.g., the intervention will have a substantial negative impact on revenue), and time-consuming (e.g., the effect may take months to materialize). Even if we can run an experiment, knowing the magnitude of the impact will significantly accelerate the product development life cycle by helping us prioritize tests and determine the appropriate traffic allocation for different treatment groups. In this setting, we should leverage observational data to quickly and cost-efficiently obtain a reliable estimate of the causal effect. Although causal inference from observational data has a long history, its adoption by data scientist in technology companies has been slow. In this paper, we rectify this by providing a brief introduction to the vast field of causal inference with a specific focus on the tools and techniques that data scientist can directly leverage. We illustrate how to apply some of these methodologies to measure the effect of contributions (e.g., post, comment, like or send private messages) on engagement metrics. Evaluating the impact of contributions on engagement through an A/B test requires encouragement design and the development of non-standard experimentation infrastructure, which can consume a tremendous amount of time and financial resources. We present multiple efficient strategies that exploit historical data to accurately estimate the contemporaneous (or instantaneous) causal effect of a user's contribution on her own and her neighbors' (i.e., the users she is connected to) subsequent visitation frequency.  We apply these tools to LinkedIn data for several million members.

\end{abstract}

%
%


%
%

\section{Introduction}
%
%


Causal inference from observational data has been an active area of research for many decades~\cite{rubin1974estimating,imbens2015causal,forastiere2016interference}.  In observational studies, the user self-selects into treatment, as opposed to randomized experiments where the experimenter controls the treatment assignment. Since we relinquish control of the assignment, the validity of our inference will depend on untestable assumptions to disentangle the treatment effect from the bias caused by self-selection~\cite{rosenbaum1984reducing,eckles2017bias}. Nevertheless, as we show, by carefully selecting features and using appropriate models we can obtain reliable estimates. 

Our work is motivated by an application at LinkedIn, a professional social-network consisting of hundreds of millions of members connected through billions of edges. A constant theme in product development at LinkedIn is understanding which features add the most value to members' experience, where the value is typically measured by the amount of time spent on LinkedIn or the visitation frequency. One such value-adding initiatives recently conceived was to help members stay better connected by encouraging them to contribute more on LinkedIn, be it public contributions (e.g., posting, commenting sharing or liking a feed) or private contributions (e.g., messaging another member). We believe that such contributions serve as a starting point for conversations and lead to more frequent visits to LinkedIn. Our primary goal is to quantify this relationship and measure the effect of a contribution on a user's subsequent visitation frequency, and the impact on their neighborhood (i.e., the users she is connected to) using historical data. Our secondary goal is to identify how the effect differs across different user segments, and determine the focus of future initiatives. 

Web-facing companies, such as Amazon, Facebook, and Google, often employ strategies that encourage a member to take an action because of the belief that this will impact a downstream business metric. Measuring the effect of an action taken by a user on a business metric can be done using experiments, but requires encouragement design \cite{imai2013experimental} (since we can not directly force users to take the desired action) and developing non-standard experimentation infrastructure \cite{eckles2016estimating}. In this setting, however, we usually have an abundance of historical data to which we can apply an observational causal method to estimate the causal effect. 

The use of observational studies extends far beyond the use cases discussed in this paper. Below we list some situations where data scientists at LinkedIn have used an observational study with tremendous business impact: 

\emph{Impossible to A/B test.} Typically this situation occurs when working with external partners or vendors. For example, updating or launching a new mobile application cannot be easily A/B tested as a third party vendor usually controls the release process. Even though both Andriod Play store and iOS Store now allow staged roll-out (where initially only a small proportion of users with auto update receive the latest version) the application owner does not know which members are eligible for the upgrade. Moreover, as soon as the application is released all new user receives the latest version and old user can manually update. Therefore, there is two-sided non-compliance (i.e., users in treatment A can take B and vice-versa) and the actual treatment labels are missing. To adequately measure the relative performance of the new version relative to the old version we have to use observational causal methods \cite{xu2016evaluating}.

\emph{ Metric definition evaluation.} At LinkedIn, there are four company-wide true north metrics. Each product area develops pillar specific metrics that are more sensitive to their innovations. A question that naturally arises is, are the pillar specific metrics good surrogates for the true north metrics? In other words, does optimizing for the pillar specific metrics necessarily lead to the optimization of true north metrics? To answer this question, we can again conduct an observational study where the treatment is increasing the pillar-specific metric, and the response is the true north metric.

\emph{ Marketing campaigns (both offline and online).} Marketing campaigns are typically hard to evaluate through standard randomized experiments. Especially if the campaign is targeting a specific geo-location, usually all the members in that area are exposed to the treatment. We can then use causal methods to analyze the relative effectiveness of the campaign \cite{brodersen2015inferring}.

In this paper, we provide an introduction to the basics of causal inference from observational studies. We discuss some of the underlying assumptions and standard methods, for some we also provide references for software implementations. We then present an approach based on linear fixed effects models to estimate the contemporaneous (or instantaneous) causal effect of a user's contribution on her own and her neighbor's subsequent visits (also known as the peer effect). This approach is particularly useful for data scientist as it covers multiple use cases. Our main methodological contribution is the insight that we can translate a problem of peer effects into one of temporal ordering. By using temporal data and a carefully designed observational study, we can reduce the self-selection bias and obtain accurate measures of the causal effect. 

The structure of the paper is as follows. In Section \ref{S:obs-study} we provide a comprehensive introduction to observational studies. In Section \ref{S:notation} we focus on the special case where we have multiple observations for the same member under different treatments. We then define the contemporaneous causal estimand, detail our assumptions and present a model for estimating the causal effect. In Section \ref{S:application} we describe our empirical application. In Section \ref{S:sim} we present a small simulation to explain the benefits of some of these approaches. Finally, in Section \ref{S:conclusion} we provide an overview of our internal platform for performing observational causal analysis. 

\section{Introduction to observational studies}\label{S:obs-study}
The main distinction between an A/B test and an observational study is that in the former we (the experimenter) control which treatment a member goes into, whereas in the latter the members decide which treatment to take \cite{rubin1974estimating}. Therefore, in a randomized experiment, we know the treatment assignment probabilities and can ensure that the treatment assignment is independent of the potential outcomes (conditional on some features). In an observational study, however, we do not know these probabilities and the outcomes are correlated with the individual features. The inherent correlation leads to systematic differences between the treatment and control groups. Naively comparing the difference in means will provide a biased estimate for the true causal effect. The goal of an observational causal analysis is to remove (or at least reduce) this bias and obtain plausible causal estimates.

\subsection{Mathematical formulation}

To ensure that it is possible for there to be a causal relationship, the intervention (or treatment) must happens before the outcome is measured. For member, $i =1,...,N$ let $W_i \in \{0,1\}$ be the treatment assignment, let $X_i$ be set of features that we have, and let $Y_i (0)$ be the outcome if member $i$ took control and let $Y_i (1)$ be the outcome if they received treatment. For example, the treatment is contributing to LinkedIn, the outcome is the number subsequent visitations, and the features are member level characteristics such as location, past visitation frequency, and job seeker status.  The potential outcomes represent two alternative realities; in one we administer control to member $i$ and observe outcome $Y_i (0)$, in the other, we administer treatment 1 and observe an outcome $Y_i (1)$. After the treatment is assigned we only observe one of these two outcomes, the other is missing. 

By writing the potential outcomes for member $i$ only as a function of the treatment assignment for that member, we are implicitly assuming that there is no interference. That is, the treatment received by other member does impact member $i$'s potential outcomes. Moreover, we are also assuming that every member is receiving the same version of the treatment. The combination of these two assumptions often referred to as the stable treatment value assumption (SUTVA) \cite{rubin1986comment}.

In a randomized experiment, we control the treatment assignment and can thus ensure the treatment assignment is independent of the potential outcomes,
\begin{equation}\label{E:unconfounded-rand}
    Pr(W_i = 1| Y_i (0), Y_i (1))=Pr(W_i = 1 ) \text { or } W_i \independent  Y_i (0), Y_i (1),
\end{equation}
where $\independent$ is the symbol for independents between two random variables. 
This is the key assumption allowing us to claim that correlation is indeed causation. It ensures that any differences we see in the outcomes between the members in treatment and control is due to the intervention as opposed to systematic differences between the two treatment groups. That is why A/B testing is so popular in industry, it allows us to obtain causal estimates of the success of new products and innovations. 

In an observational study, we do not control the treatment assignment; therefore, we can no longer guarantee that (\ref{E:unconfounded-rand}) holds. However, in many examples, it is reasonable to assume that decision to favor one treatment over an alternative is predominantly determined by a member's past behaviors or features. In that case, a weaker version of (\ref{E:unconfounded-rand}) may hold asserting that the treatment assignment only depends on a set of observed features, 
\begin{equation}\label{E:unconfounded}
    Pr(W_i = 1| Y_i (0), Y_i (1), X_i )=Pr(W_i = 1 |X_i) \text { or } W_i \independent  Y_i (0), Y_i (1) | X_i,
\end{equation}
Under this assumption, within groups defined by $X_i$, it is as if we had conducted a randomized experiment. Traditionally, the causal analysis of observational studies is split into three parts.
\begin{table}[t]
    \centering
    \caption{R packages for causal inference}
    \begin{tabular}{c c c}
    \hline \hline
        Matching &  Weighting &	Temporal \\ \hline
         \texttt{FLAME}\cite{roy2017flame} &	\texttt{CausalGAM} \cite{glynn2010causalgam} & \texttt{Causalimpact} \cite{brodersen2015inferring} \\
         \texttt{MatchIt} \cite{ho2011matchit} & \texttt{IPW}  \cite{van2011ipw} & \texttt{wfe} \cite{kim2017weighted} \\
         \texttt{CEM} \cite {iacus2009cem} & \texttt{sbw} \cite{zubizarreta2015stable} & \texttt{lfe}\cite{gaure2013lfe} \\
        \hline
    \end{tabular}
    \label{tab:r-packages}
\end{table}

\subsubsection{Step 1: The design}
In this step, we attempt to remove the systematic differences between the treatment groups. Intuitively, we want to find a set of units that are comparable, meaning that they have the same features and therefore resemble a randomized experiment. Typically, this step is performed by only utilizing the features, because looking at the outcome can introduce unintended biases \cite{imbens2015causal}. 

A simple starting point when $X_i$ are all categorical is to perform an exact match (i.e., group by all $X_i$ and only keep the buckets that have at least one member in treatment and control) \cite{rubin2006matched}. If some of the features are continuous, then we discretize them and perform a coarsened exact match \cite{iacus2009cem}. There are many alternative methods for implementing the matching; see Table \ref{tab:r-packages} for more examples.

As the number of features increases, it becomes desirable to perform some dimension reduction. We usually do this through the propensity score
\begin{equation}
    e(X_i ) = Pr(W_i = 1| X_i ),
\end{equation}
a sufficient one dimensional summary of all of the features \cite{rosenbaum1983central}. The propensity score is a balancing score, ensure that $W_i \independent X_i | e(X_i)$, which implies that $W_i \independent (Y_i(0), Y_i(1)|e(X_i)$.
Therefore, if two members have the same propensity score (meaning that they have equal probability of taking treatment), but one took treatment whereas the other took control, then it is as if these two members were part of a standard A/B test where the probability of being in treatment 1 is $e(X_i )$. In practice, we do not know the propensity score; instead, we estimate it using, for example, a logistic regression. 

We can use the propensity score in many different ways to remove the inherent systematic differences. Examples include, weighting, matching, and stratification, see Table \ref{tab:r-packages} for references of \texttt{R} \cite{R-Core-Team:2013aa} packages. Weighting attempts to recreate a balanced group by appropriate weighting the units so that on average they resemble a randomized experiment \cite{rosenbaum1983central}. Matching identifies one (or more) control units that closely resembles each treatment unit \cite{rubin2006matched}. Stratification, groups members by their propensity score in such a way that the propensity score for a given group is roughly constant across the members in treatment and control \cite{imbens2015causal}. 

If this step is carried out correctly, this should remove any systematic variation due to the measured features, ensuring that the remaining units resemble a randomize experiment. 

\emph{Checks and balances}

After performing the design stage of an observational study, it is critical to assess the balance that was achieved. The simplest way to do this is to look at the standardized difference in means for the features between the members in the different treatment arms \cite{imbens2015causal}.

\subsubsection{Step 2: The analysis}
After a successful design stage, we can analyze the balanced cohort like a standard randomized experiment. For example, we could use a regression analysis on the balanced set.\footnote{The estimator obtained from combining weighting and regression methods is often referred to as doubly robust \cite{imbens2015causal}.} This requires the assumption that there are no unmeasured features. For example, if we are trying to measure the effect of contributions, then a potential unmeasured feature is a person inherent likelihood of contributing. However, if the unobserved features are positively correlated with the observed ones, then the design stage should have removed a substantial amount of the bias. 

\subsubsection{Step 3: Diagnostic tools and sensitivity analysis}
After performing the analysis, it is paramount to use diagnostic tools to detect divergence away from the underlying assumptions. Below we suggest two simple methods which we have found to yield good results in practice.

\emph{Backward causality:} Achieving a good balance ensures that the potential outcomes do not contain any information about the treatment assignment. Hence, if we run a regression for  $Pr(W_i = 1| Y_i, X_i )$ on the balanced set, we should see that the coefficient for $Y_i$ is close 0. Otherwise, we were unable to account for the systematic variation between the two treatment groups. In practice, we diagnose backward causality by rerunning the analysis procedure treating the outcome as the treatment assignment and the treatment assignment as the outcome. 

\emph{A/A testing:} Typically, the outcome of interest in observational studies is a member level metric which is measured over time. Therefore, we usually have a measure of the metric before the intervention occurred. Using the historical metrics as our primary outcome and applying our chosen analysis method to this outcome should yield a causal estimate of 0. If the result differs from zero, then we should return to step 1 and repeat the procedure. 

After performing diagnostic tools we can use sensitive analysis methods\footnote{We define sensitivity analysis as a comparison of multiple models on the substantive conclusions attained from our initial inference.} to assess the robustness of our results \cite{imbens2015causal}.

\section{Observational studies with temporal data} \label{S:notation}
Technology companies often record how members interact with their platform over time. We can use this temporal data to improve the quality of some of our causal estimates. Returning to our motivating example, we can track member contribution status as well as their subsequent visitation frequency. This allows us to use each member as their own control. Below, we provide the details on how to extract causal estimates from temporal data. 

\subsection{Set up}
Assume we track $i=1,\dots, N$, users' activities over time, $t = 1, \dots, T$, with $T\ge 2$. For user $i$ at time $t$, let $Y_{i,t}$ be the outcome, let $\mathbf{X}_{i,t}$ be a vector of (time varying) covariates, and let $W_{i,t}$ be the treatment status. In our application $Y_{i,t}$ is the visitation frequency and $W_{i,t}$ is the contribution status ($W_{i,t}=1$ if the user contributes and 0 otherwise). Prior to commencing the study, we also measure a set of baseline covariates $\mathbf{X}_{i,0}$ that are constant throughout. To distinguish between random variables and their realizations, we will use lower case letters for the latter. Finally, we denote the cumulative history using ``$:$'' notation, for example, the ``treatment path'' is $W_{i,1:t} = (W_{i,1},\dots, W_{i,t})$ \cite{bojinov2017}.

\begin{figure}[t]
    \centering
    \includegraphics[width=0.5\textwidth]{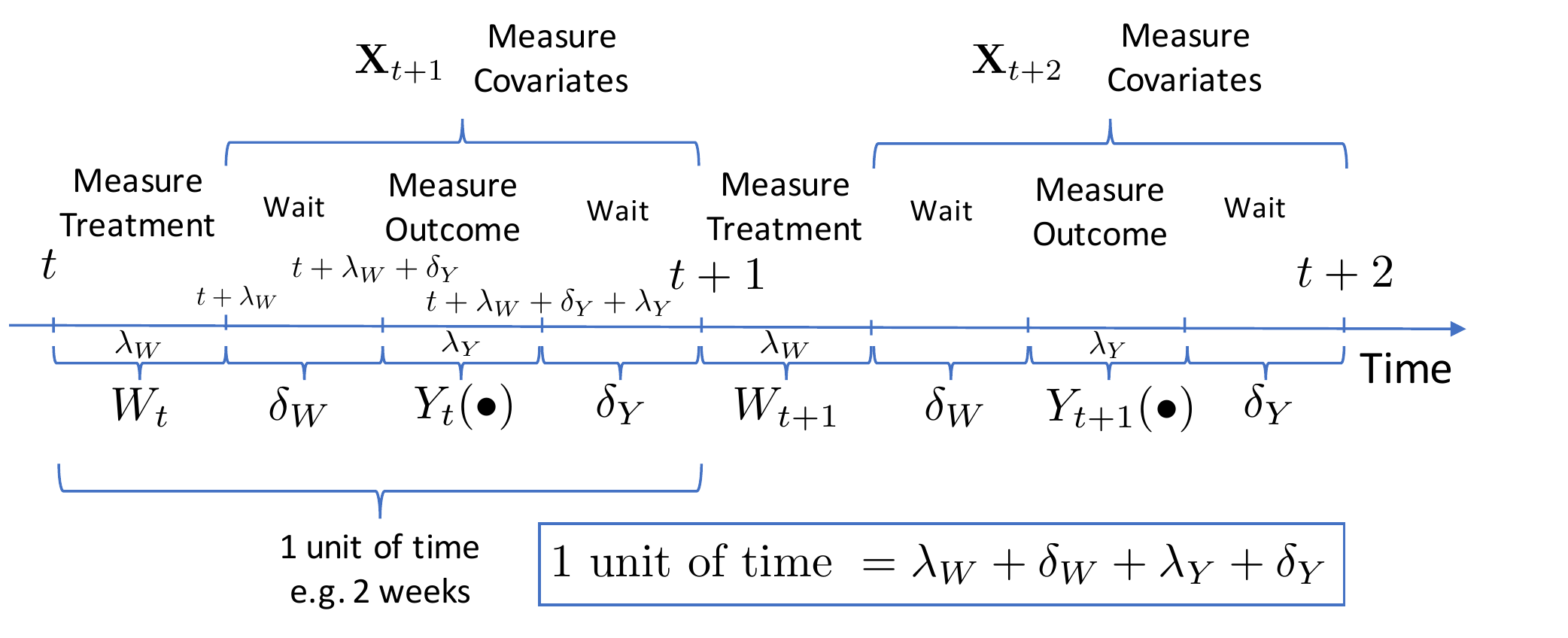}
    \caption{A visualization of the data measurement process. We first measure the features, then the treatment assignment, and finally the outcome. }
    \label{fig:time}
\end{figure}

To ensure that we can obtain a causal interpretation, we first measure the covariates, then the treatment assignment (contribution status) and finally the outcome (visitation frequency). Each of the measurements is taken over some interval of time, for example, the contribution status is measured between time $t$ and $t+\lambda_W$. Figure \ref{fig:time} provides a visualization of the process.

For now, we assume that there is no interference between users and there are no hidden version of treatment, this is known as the temporal stable unit treatment value assumption (TSUTVA)\footnote{TSUTVA is a generalizes of the traditional stable unit treatment value assumption~\cite{rubin1980randomization}.} \cite{bojinov2017}. To define causal effects, we use the potential outcome framework~\citep{rubin1974estimating, robins1986new, bojinov2017}; under this formulation, the outcome is a function of the full treatment path, denoted by $Y_{i,t}(W_{i,1:t})$. Our notation makes it clear that visitation frequency at time $t$ is a function of the full contribution history, this allows contributions at time $s\le t$ to impact on the number of visits at time $t$. Since the covariates, at time $t$, are measured after the $(t-1)^\text{th}$ contribution, they too are affected by the treatment path, that is $X_{i,t+1}(W_{i,1:t})$.

We define user-level causal effects as any contrast of potential outcomes at a fixed time point \cite{rubin1990formal,bojinov2017}. In our application, we are primarily interested in a special case known as contemporaneous causal effect \cite{bojinov2017}. The contemporaneous (or instantaneous) effect, measures the average effect of contributing, at time $t$, on visitation frequency, at time $t$. 
Formally, the contemporaneous effect for members with covariate history $S_t$, baseline covariate in $A$, and treatment path $w_{i,1:t-1}$ is,
\begin{align}\label{E:causal-effects-one-path}
    \tau_{t}(1,0 &| A, S_{t}, w_{i,1:t-1})) = \E[Y_{i,t}(w_{i,1:t-1}, 1)-Y_{i,t}(w_{i,1:t-1}, 0)  | \notag \\ &\mathbf{\mathbf{X}}_{i,1:t}(w_{i,1:t-1})\in S_{t}, \mathbf{X}_{i,0} \in A, W_{i,1:t-1} = w_{i,1:t-1}],
\end{align}
where the expectation is taken with respect to a super population, under the assumption that the users participating in the study were obtained using simple random sample \cite{imbens2015causal}. As an example, let $A$ be the set of users living in the USA, $S_t$ be the set of users that have more than 500 neighbors at time $t$, and $w_{i,1:t-1} = (0,\dots,0)$. Then, \eqref{E:causal-effects-one-path} measures the average effect of the first contribution taking place at time $t$ for users based in the USA with more than 500 neighbors. To obtain an estimate for the whole population we can set $S_t = Im(\mathbf{X}_{i,1:t})$, and $A = Im(\mathbf{X}_{i,0})$, where $Im(f)$ is the image of a function $f$ \cite{sobel2012does}. 

Usually, we are not interested in the effect of one particular treatment path, but instead, want to average over the historical treatment paths to obtain the average contemporaneous effect,
\begin{equation}\label{E:causal-effects}
    \tau_{t}(1,0 | A, S_{t})) = \E_W [\tau_{t}(1,0| A, S_{t}, W))],
\end{equation}
where the expectation is taken over the random treatment path. 
Continuing with our earlier example, \eqref{E:causal-effects} measures the average effect of contributing at time $t$ for users living in the USA with more than 500 neighbors. Since the start time of our observational study is arbitrary, we expect this effect to be constant (or at least approximately constant) through time. Hence, we can drop the subscript $t$ from the causal effect.

\subsection{Inference}\label{S:inference}

One way we can estimate the contemporaneous causal effect is to take a parametric approach and assume a linear fixed effects (FE) model for the outcome,
\begin{equation}\label{E:FE}
    Y_{i,t}(w_{i,1:t}) = \beta^\top \vec{\mathbf{X}}_{i,t} + \tau^{(F)}(\mathbf{X}_{i,0})w_{i,t} + U_i + V_t+ \varepsilon_{i,t}(w_{i,1:t}),
\end{equation}
where $\vec{\mathbf{X}}_{i,t} = (1, \mathbf{X}_{i,t}(w_{i,1:t-1}))^\top$, $\tau^{(F)}(\mathbf{X}_{i,0})$ is the treatment effect (which is  potentially a function of baseline covariates), $U_i$ is a user level fixed effect, $V_t$ is a time effect, $\varepsilon_{i,t}(w_{i,1:t})$ is a potential error term, $\beta$ is a vector of unknown parameters, and the superscript $\top$ denotes the transpose of a vector \cite{sobel2012does}. We can interpret the fixed effect $U_i$ as user $i$'s baseline visitation frequency, and we can interpret $V_t$ as a controlling for any unmeasured seasonality effect. 

The advantage of this model is that it allows us to exploit the multiple observations for the same user over time. The user-level fixed effect accounts for any user specific variation that is not captured by the observed covariates; in a sense, each user acts as their own control. Accounting for unobserved features is not possible in general one-time point analysis as there is no information about the unobserved potential outcome without having multiple observations for the same user.

For the FE estimator $\hat\tau_t^{(F)}$ to estimate the contemporaneous causal effect \eqref{E:causal-effects}, we need to impose further assumptions which generally cannot be falsified using the observed data. In what follows we details these assumptions and provide some justification for why we believe they hold in our application, see \citet*{sobel2012does} for an indepth discussion of each of these. Of course, the linear model may not hold exactly, but it can still be useful in obtaining reliable estimates. 

We are going to assume that the vector of covariates does not contain lagged version of the outcome, this is often called the "strictly exogenous" assumption \cite{chamberlain1982multivariate}, and it ensures that the error term is on average centered around 0,
\begin{equation}\label{E:se}
    \E[\varepsilon_{i,t}(w_{i,1:t})| \mathbf{X}_{i,0:T}, W_{i,1:T}, U_i,V_T] = 0,
\end{equation}
for all $t$.

We further assume that, given the observed covariates, unobserved temporal effect, and the unobserved user effect, the contribution at time $t$ does not depend on future outcomes and covariates,
\begin{align}\label{E:Assump1}
    \{\mathbf{X}_{i,s}(w_{i,1:s-1}), &Y_{i,s}(w_{1:s-1})\}_{s=t+1}^T, Y_{i,t}(w_{1:t}) \independent W_{i,t} | \\
     &\mathbf{X}_{i,0:t}, W_{i,1:t-1} = w_{i,1:t-1}, U_i,V_{1:t}.\notag
\end{align}
Assumption \eqref{E:Assump1}, requires that a user's decision to contribute does not depend on her future visitation frequencies and covariates, given the observed history. This assumption seem reasonable in our application, as users rarely plan their current contributions based on their expected future actions.

Our final assumption is that future treatments are independent of the outcome at time $t$, conditional on all past and future covariates, the unobserved temporal effect, and the unobserved user level effect, 
\begin{equation}\label{E:Assump2}
    Y_{i,t}(w_{1:t}) \independent W_{i,t+1:T} | \mathbf{X}_{i,0:T}, W_{i,1:t} = w_{i,1:t}, U_i,V_T.
\end{equation}
Assumption \eqref{E:Assump2}, requires that a user's visitation frequencies do not predict her future decisions to contribute, given the past and future covariates. We can make this assumption more plausible, by increasing the time between measuring the outcome and recording the subsequent treatment assignment. 

Under Assumptions (\ref{E:Assump1}) and (\ref{E:Assump2}), the estimates of the causal effect from a strict exogeneity fixed effects model (\ref{E:FE}) are equal to the contemporaneous causal effects given in (\ref{E:causal-effects}) \cite{sobel2012does}. 

 In our application, we believe that there is no (or at least very little) seasonality effect. This means that we can drop $V_T$ from our analysis without impacting our ability to estimate the causal effect. Moreover, removing the fixed effect for time leads the FE estimator to be mathematically equivalent to a matched sample estimator whereby each user's missing potential outcomes are estimated using weighted past, and future observed outcomes \cite{imai2012use}.

\subsection{Weighted fixed effect model}

In the single time point observational studies literature, the propensity score (the probability of a user contributing, conditional on covariates) is often used to adjust for covariate imbalances \cite{rosenbaum1983central}. We can do this through inverse propensity score weighting or propensity score stratification. Alternatively, we can also use the propensity score to improve the efficiency and robustness of a regression estimator by using a ``doubly robust'' estimator, that combines weighting and regression analysis. An extension to FE incorporating weights, referred to as the weighted FE model, was proposed in \citet*{imai2012use}. They suggested that at each point in time, the observation should be weight by the estimated propensity score. In our application, we consider both weighted and unweighted FE models. 

\subsection{Handling no-show users}

At each time point $t$, the treatment assignment is determined over the interval $(t, t+\lambda_W)$. Some users, however, are not online during this period. If we consider users that did not visit the same as the users that visit but do not contribute, then we are estimating the effect of contributing versus not contributing or visiting. Of course, this effect should, on average, be larger, but it may lead us to make incorrect business decisions as we are not estimating the right estimand.

One simple solution is only to include users that were active during every treatment assignment period. Two drawbacks of this approach are that we will inadvertently limit our analysis to heavy users, and reduce the sample size. Both of these issues will make it harder to generalize our results to the whole population of interest. 

An alternative approach is to consider the intervention as having three treatments: contributing, visiting without contributing, and not visiting. We can then directly contrast contributing versus visiting without contributing. The results from the previous section readily extend to this situation, with only minor notational changes, we can thus use the fixed effects model from (\ref{E:FE}) to estimate the causal effect of interest. 

\subsection{Alternative inference strategies and their assumptions}

An alternative approach to causal inference from temporal data under a slightly different set of assumptions has been extensively studied in the biostatistics literature \cite{robins1986new,robins2000marginal}. Their key identifying assumption is a generalization of the ignorability assumption \cite{rosenbaum1983central} often referred to as sequential ignorability \cite{robins1986new}. Sequential ignorability is different to Assumptions \eqref{E:Assump1} and \eqref{E:Assump2}, and requires that the treatment assignment at time $t$ is conditionally independent of future potential outcomes given the whole observed history. Inference in this setting for example, can be carried out using marginal structural models that model the marginal mean of the potential outcomes $E[Y_{i,t}(w_{i,1:t})]$ and use inverse probability weighting (IPW) to estimate causal effects \cite{robins2000marginal}. It is still an open problem to derive estimators that are consistent when either sequential ignorability or Assumptions \ref{E:Assump1} and \ref{E:Assump2} hold. The challenge of these methods is their inability to scale to large datasets.

\subsection{Measuring spillover effect}\label{S:peer-effect}

When measuring the effect of contributions on subsequent visits, it seems likely that for some types of contributions the no interference assumption does not hold. For example, if a user receives a direct message then her is likely to respond - this means that a user's contribution is affecting his neighbor's future visits. An estimator for the causal effect in \eqref{E:causal-effects} can be biased when there is interference \cite{eckles2017bias,toulis2013estimation,forastiere2016interference}. However, if we include neighborhood covariates in the regression, then \eqref{E:causal-effects} and the spillover effect (defined below) can be identified, within groups of emembers with the same netwrok structure \cite{forastiere2016interference}. 

Assume that our members are connected on network $\mathcal{G}$ where the nodes represent users, and the edges represent connections. Each unit $i$ has a connection neighborhood, denoted by $\mathcal{N}_i$. Assume that the potential outcomes can be indexed by a user's treatment path and a function of his neighborhoods' treatment path. That is, a user's visitation frequency is a function of her own contributions and an average of the contributions in her neighborhood. Mathematically, $Y_{i,t}(W_{i,1:t}, Z_{i,1:t})$ where $Z_{i,t} = \log_2 \sum_{j\in \mathcal{N}_i} W_{i,t}$ is the base two logarithm\footnote{We used the base two logarithm to allow for easy interpretation of the results.} of the sum of contribution statuses of member $i$'s neighborhood\footnote{Units that have no connections will be dropped from the analysis, this ensures that the potential outcomes are well defined for all members.}. The neighborhood treatment $Z_{i,t}$ is a deterministic function of $W_{j,t}$, $j\in \mathcal{N}_i$, that takes values in some finite set $\mathcal{Z}_k$ where $k$ is the number of members that member $i$ is connected to, denoted by $|\mathcal{N}_i|$. 

In our application, it seems reasonable to assume that the neighborhood contributions only have an immediate impact on the outcome. Specifically, we evoke the ``$m=0$'' causal structure assumption \cite{bojinov2017},

\begin{equation}
    Y_{i,t}(W_{i,1:t}, Z_{i,1:t}) = Y_{i,t}(W_{i,1:t}, Z_{i,1:t}') \text{ whenever } Z_{i,t} = Z_{i,t}',
\end{equation}
We can now write the potential outcomes as $Y_{i,t}(W_{i,1:t}, Z_{i,t})$. 

We define the (contemporaneous) spillover effect as the average effect of an increase of $\nu$ contributions in a user's neighborhood at time $t$ on her visitation frequency, at time $t$. The spillover effect is formally defined as,
\begin{align}
    \delta_t(z,\nu|&A, S_t, w_{i,1:t}) = \E[Y_{i,t}(w_{i,1:t},z+\nu) -\notag Y_{i,t}(w_{i,1:t},  z) | \notag \\
    &\mathbf{\mathbf{X}}_{i,1:t}(w_{i,1:t-1})\in S_{t}, \mathbf{X}_{i,0} \in A, W_{i,1:t-1} = w_{i,1:t-1}],
\end{align}
where we implicitly condition on the set of units for whom $z \in \mathcal{Z}_k$. Similarly to the main effect, we also average over the historic treatment path to obtain $\delta_t(z,\nu|A, S_t)  = \E_W[\delta_t(z,\nu|A, S_t, W) ]$. 

To estimate the spillover effect, we will use a similar strategy to the one described in Section \ref{S:inference}. In particular, we will assume a linear fixed effects model except now we will include an additional parameter for the spillover effect. To identify the causal effect, we will also include the summary statistics of the neighborhood covariates \cite{forastiere2016interference}. Under the analogous set of assumptions the FE estimator will estimate the spillover effect \cite{sobel2012does}.

\section{Applications}\label{S:application}

We now apply the methods introduced in the previous sections to estimate both the effect of contributing on a member's visitation frequency in the subsequent week as well as the impact due to her neighbor's activities. To allow for better interpretability, we separately considered the impact of public and private contributions. 

A neighborhood's public contributions can impact a user's visitation frequency through three mechanisms. First, the member can receive a notification of a response to her public contributions; this can take the form of a comment, like or share. Second, the member can receive a notification of a public post from her neighborhood; typically these are viral posts. Third, an increase in the general quality and quantity of her feed's content pool can motivate members to increase their base visitation frequency.

Private contributions are conversations that happen within private channels that are only visible to members of the channel. We will only focus on direct messaging between members as that is by far the most common type of private contribution on the LinkedIn platform.

For both effects, we also study the impact on different subpopulations (for example, highly engaged members and new members) as understanding these is crucial for determining resources allocation. 
\subsection{Data collection}

We measured the contribution status over 1 day ($\lambda_W=1 \text{ day}$), and the visitation frequency over the subsequent 7 days ($\lambda_Y =7 \text{ day}$). There was no gap between the two measurements as we are interested in the short term impact of contributing. We measured the contribution status only for one day because the impact of a contribution diminishes over time. From historical metrics and experiments, we know that weekend traffic is much lower than weekday traffic, and weekend visitors also behave and respond to experiments differently from weekday users. By changing the day of the week for which the contribution was measured, we were able to average over any such differences. 

The features \(\mathbf{X}_{i,0}\) contain user-level information (such as country, industry, connection count, historical engagement) and were collected over the month before the start of the study. The time-varying covariates $\mathbf{X}_{i,t}$ contain updated versions of the engagement history and connection count.  

\subsubsection{Main Effect}

To estimate the main effect, we label member \(i\) at time \(t\) to have treatment 1 if she did not visit, 2 if she visited but did not contribute and 3 if she visited and contributed.  To perform subgroup analysis based on member life cycle (MLC): new members (known as onboarding), monthly users, weekly users and daily users. For the subgroup analysis we generated treatment labels by considering the interaction of contribution and member life cycle. 

\subsubsection{Spillover Effect}

In addition to the features described above, we also include summary statistics of their neighbors' covariates in \(\mathbf{X}_{i,t}\) for modeling spillover effect. The model also include an indicator for the self-initiated contributions during the treatment measurement period (we separated out the self-initiated and respond-form contributions as the latter could be attributed as an outcome of their neighbors' contributing actions). To measure and compare effects of different segments of members, we conducted separate analysis that limits the population to users who were in the same engagement level throughout the entire analysis period.

\subsection{Analyses and results}

We explored various methods includes naive correlation, weighting methods, regression ~\cite{rosenbaum1983central}, doubly robust, FE (\ref{E:FE}) and weighted FE models. 

We could only use one data for one period to compute the correlation and doubly robust estimates. The results were reasonably stable across the different time periods; hence we only show one of them. The FE model  can be estimated directly, for example, by using the \texttt{lfe} \citep{gaure2013lfe} package in \texttt{R} \citep{R-Core-Team:2013aa} . For the weighted FE, we must model the probability of a member contributing as a function of \(\mathbf{X}_{i,t}\).  

\subsubsection{Main Effect}


\begin{figure}[t]
    \centering
    \includegraphics[width=0.54\textwidth,trim={3cm 1cm 1cm 1cm}, clip]{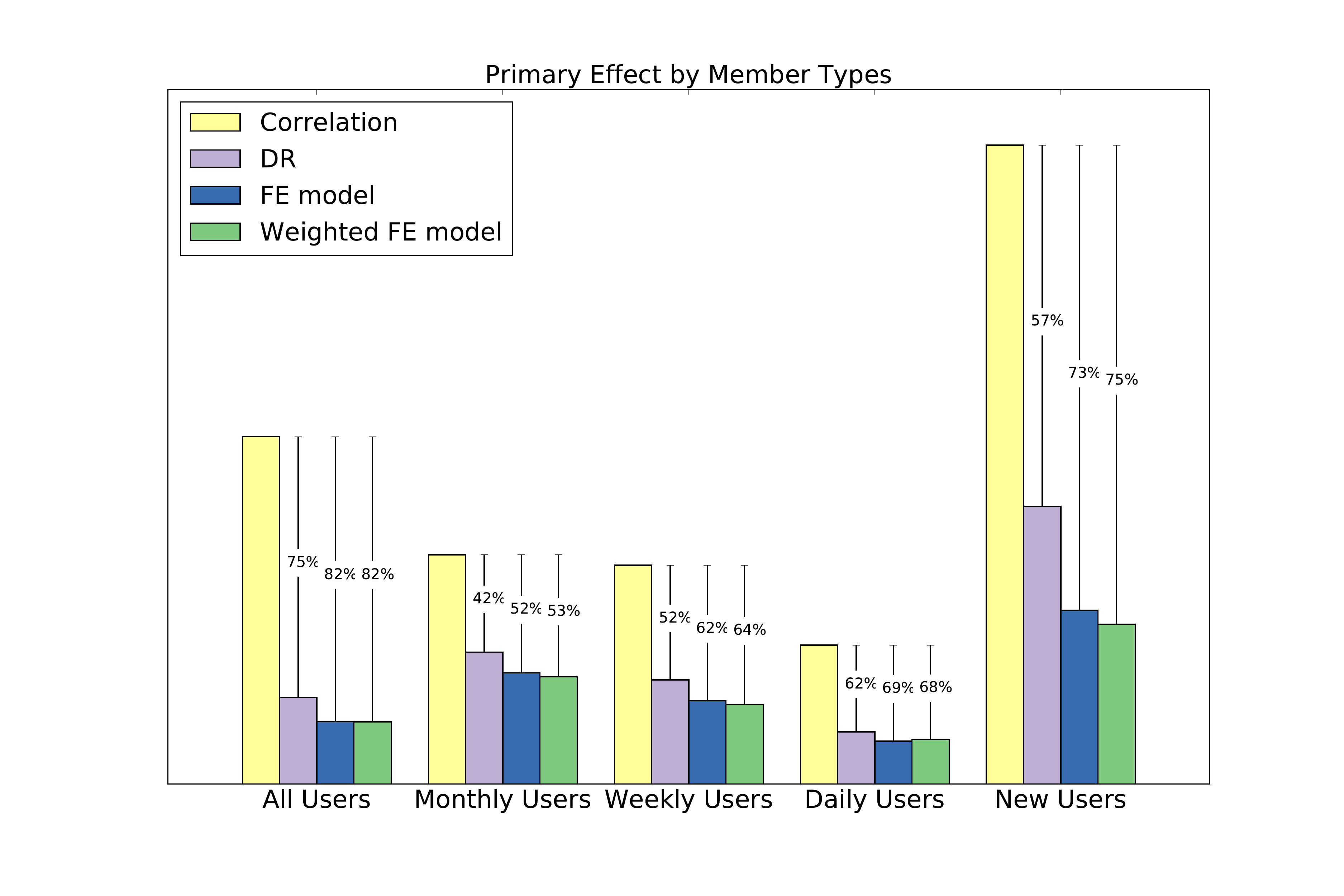}
    \caption{The contemporaneous primary effect estimate, split by the different member life cycles, using four different estimation approaches: correlation (yellow),  single time doubly robust (purple), fixed effects (blue), and weighted fixed effects (green).}
    \label{fig:main_mlc}
\end{figure}

From Figure \ref{fig:main_mlc}, we can see that the weighted fixed effect model almost always provides the smallest estimate of the causal effect. There is little difference between the estimates obtained from the FE model and the weighted FE model; this is reassuring as it means that our estimates are somewhat stable. In this setting, observational studies tend to overestimate the effect; we can, thus, interpret the reduction as removal of bias. Moreover, estimates from later experiments corroborated these results. Therefore, the weighted FE model reduces ~\(70\%\) more bias than a sole correlation analysis. We also observe that the weighted FE model almost always reduces ~\(10\%\) more bias than doubly robust. The difference is starkest when analyzing the effect of contribution on new users, from whom we can collect little historical data to account for self-selection bias in the standard doubly robust approach.

From this analysis, we can conclude that the less engaged users (new and weekly users) see the most significant gain from contributing. If we had used a purely correlational analysis, we would have falsely concluded that the monthly and weekly users have similar causal effects. Moreover, we would have considerably overestimated the effect for new members if we used either correlation or doubly robust methods. 

Whenever we use a data-driven approach to allocating resources, it is critical that we accurately measure the effect of a policy. When the estimate is several times larger than it is, we will invest more resources than is beneficial to the company. 
The results from our analysis, combined with the size of each member class, led us to reprioritize for which user type we allocated the most resources in driving contributions. The results of this analysis also helped us design further, better, experiments.

\begin{figure}[t]
\includegraphics[width=0.54\textwidth,trim={3cm 1cm 1cm 1cm}, clip]{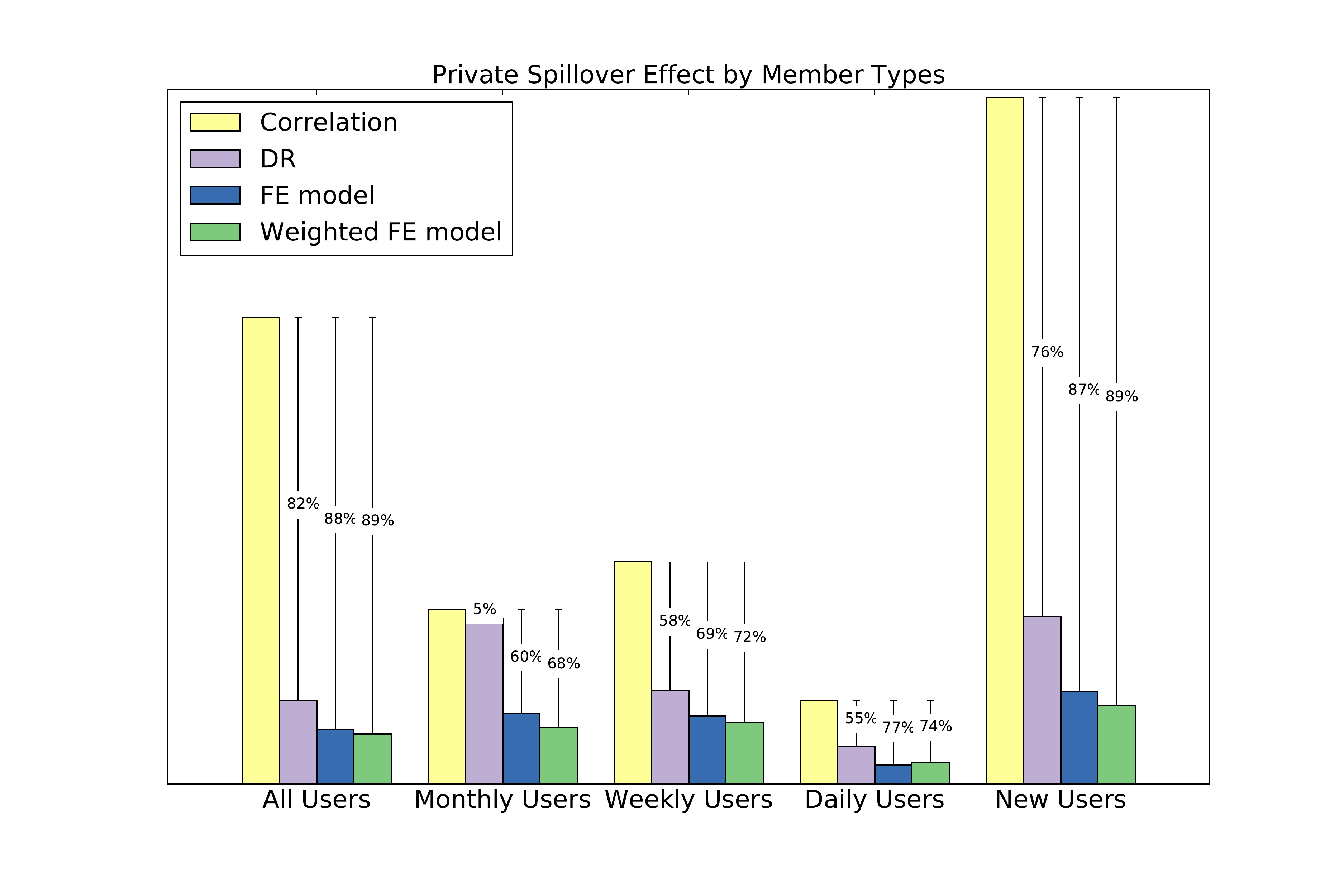}
    \centering
    \caption{The contemporaneous spillover causal effect estimates of private contributions, split by the different member life cycles, using four different estimation approaches: correlation (yellow),  single time doubly robust (purple), fixed effects (blue), and weighted fixed effects (green).}
    \label{fig:peer_mlc}
\end{figure}

\subsubsection{Spillover Effect}

For public contributions, we define the treatment as the base two logarithm of the sum of the specific type of contributions. The strategy of measuring private contribution spillover effect is somewhat different; member $i$ at time $t$ is labeled to have treatment 1 if she received at least one private contributions and 0 if she did not receive any. The baseline comparison estimates are computed using correlation and linear regression model with covariates. 

\begin{figure}[t]    
\centering
\includegraphics[width=0.54\textwidth,trim={3cm 1cm 1cm 1cm}, clip]{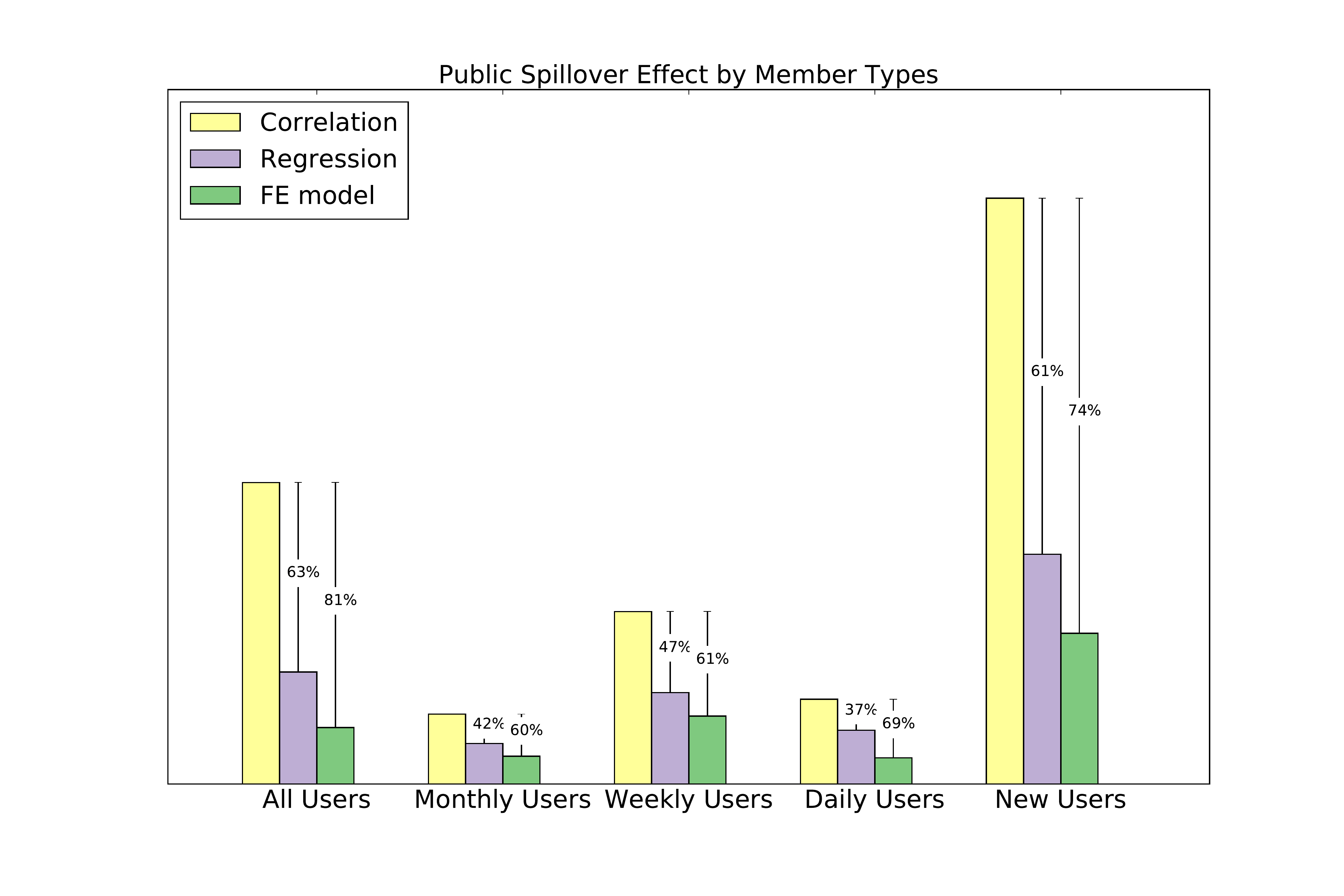}
    \caption{The contemporaneous spillover causal effect estimates of public contributions, split by the different member life cycles, using four different estimation approaches: correlation (yellow), single time regression (purple), fixed effects (green).}
    \label{fig:peer_mlc1}
\end{figure}

From Figure \ref{fig:peer_mlc}, we also observed that the ranking of opportunities based on raw correlations is different from the ordering obtained using the doubly robust and FE estimated. The change in ranks, lead us to a different allocation of resources and prioritization of tasks. 

Through the study, we estimate the magnitude of the spillover effect in comparison with the main effect for the first time. The analysis suggests that the total peer effects from contributions on members' engagement can be even higher than the main effect. These results revealed a tremendous opportunity for driving engaged members to contribute, activating their peers and leading to an overall more interactive ecosystem. Our analysis facilitates a company-wide strategy shift towards building active communities to lift engagements.

\section{Simulation study}\label{S:sim}

To illustrate the benefits of using a causal method as opposed to a naive corrolational analysis, we performed a small simulation study.
We considered a correlation, inverse probability weighting, regression, doubly robust, unweighted fixed effect and weighted fixed effect strategies for causal inference. To make the simulation more realistic we used real LinkedIn member data to inform the covariates and network structure. We used different generating models in the simulation of the primary effect and spill-over effect as different assumptions are made in the analysis stage.

\subsection{Simulation models for the contemporaneous effect}
\label{section:sim_main_effect}

Consider a social network with $N$ members, where each member $i$ has observed covariates $X_{i,t}$, an unobserved covariate $U_{i,t}$ that is correlated with the member's propensity to contribute, and another unobserved covariate $V_{i,t}$ that is correlated with the likelihood to visit  at each time $t = 1,2,..., T$. The probability of member $i$ contributes at time $t$ is generated from
\[
p_{i,t} = f\left(\alpha_t + \beta_t^T X_{i,t} +\gamma_t U_{i,t}\right),
\]
where 
 $f:\mathbb{R}\to [0,1]$ is a logistic function,
 $\alpha_t$ is the average proportion of members that contribute at $t$,
 $\beta_t$ is the effect of the covariates on contributing, and
 $\gamma_t$ is the effect of the unobserved likelihood to contribute variable.
We assume the treatment assignment $W_{i,t}$ is generated from a Bernoulli distribution $Bern (p_{i,t})$.
The outcome $Y_{i,t}$ follows a normal distribution
\[
Y_{i,t} \sim N(\delta_t^T X_{i,t} + \tau_t W_{i,t} + \xi_t V_{i,t},\ \sigma_t^2),
\]
where
$\delta_t$ is the effect of the covariates on the outcome at time $t$,
$\tau_t$ is the treatment effect,
$\xi_t$ is the effect of the unobserved likelihood to visit Linkedin on the outcome, and
$\sigma_t^2$ is the variation in the outcome. Notice that the assumptions for all of the models are violated whenever any of the parameters or the unobserved covariates vary over time. 

\begin{table*}[htb]
\caption{Simulations study results. All methods perform better than correlation, the fixed effect and weighted fixed effects consistently are closest to the truth.  } 
\centering 
\begin{tabular}{c c c c c c c c} 
\hline\hline 
Scenario & Unobserved confounders & True Effect & Correlation & IPW & Doubly Robust & FE & Weighted FE \\
[0.1ex] 
\hline 
Contemporaneous & Time-invariant  & 5 & -14.95 & 6.00 & 6.20 & 4.98 & 4.99 \\ 
& Time-varying & 5 & -14.21 &  7.95 & 6.94 & 5.73 & 6.49 \\
Public & Time-invariant & 1 & 6.93 & - & 6.49 & 1.00 & -  \\ 
&Time-varying  & 1 & 1.34 &- & 1.15 & 0.99 &-  \\
Private &Time-invariant & 10 & -72.53 & 20.37 & 19.56 & 10.00 & 9.85 \\ 
& Time-varying  & 1 & -80.09 & 81.85 & 19.27 & 0.97 &  1.37 \\
[1ex] 
\hline 
\end{tabular}
\label{table:main_sim} 
\end{table*}

\subsection{Simulation models for the spillover effect}\label{section:sim_peer_effect}

Now we assume that each member is connected to a set of other members, denoted by $\mathcal{N}_i$. In this setting, we use the same model as the one introduced in Section~\ref{section:sim_main_effect}, except that the probability of member $i$ contributes and the outcome are additionally affected by its neighbors. Formally, we assume
\begin{align*}
    p_{i,t} &= f\left(\alpha_t + \beta_t^{T} X_{i,t} + \nu_t\sum_{j\in \mathcal{N}_i} X_{j,t} +\gamma_t U_{i,t}\right) \\
    q_{i,t} &= f\left(\alpha_t' + \beta_t^{T'} X_{i,t} + \nu_t'\sum_{j\in \mathcal{N}_i} X_{j,t} +\gamma_t' U_{i,t}\right) \\
    Y_{i,t} &\sim N\left(\delta_t^{T} X_{i,t} + \tau_t W_{i,t} + \tau_t' g\left(\sum_{j\in \mathcal{N}_i} W_{j,t}\right) + \xi_t V_{i,t},\ \sigma_t^2\right),
\end{align*}
where $p_{i,t}$ and $q_{i,t}$ are a member's and her peer's probability to contribute. Function $g$ is the logarithmic function for public contributions and the indicator function for private contributions. $\tau_t$ is the primary treatment effect and $\tau_t'$ is the spillover effect. 
Similarly to the model for the contemporaneous effect, the model is misspecified whenever any of the parameter or the unobserved covariates vary over time.

We considered two scenarios corresponding to both the time-invariant $U,V$, and time-varying $U,V$. 

\subsection{Simulation Results}
\label{section:sim_results}

The results in Table \ref{table:main_sim} show that observational causal inference methods can estimate the impacts far more accurate than the naive correlation approach, even when the modeling assumption are invalid. While the correlation suggests a negative relationship between the treatment and outcome variable, all three causal inference techniques provide us with much more reasonable estimates (positive and close to the truth) for the contemporaneous effect. As expected, the spillover effect simulation results in Table \ref{table:main_sim}  further show the robustness of FE models towards addressing omitted covariates bias. In the case that if exists a large unobserved time-invariant confounder, the regression without member-level fixed effect would suggest an estimate that is more than five times greater than the real value. 

Also, based on all Tables, the FE and weighted FE models can outperform the standard inverse propensity weighting (IPW) and doubly-robust regarding the biases reduction. Especially when the unobserved confounders are time-invariant, the FE models almost perfectly quantify the true effect. For public contributions, both the correlation and regression methods perform better when $U$ and $V$ vary with time because this reduces the user level systematic variation. 

From this simulation study, we can conclude that our modeling approach is somewhat robust to minor model misspecification and leads to a reliable conclusion.


\section{Platform for causal inference}\label{S:conclusion}
\begin{figure}[htb]
    \centering
    \includegraphics[trim=0cm 0cm 0cm 1.5cm,clip=true,width=0.5\textwidth]{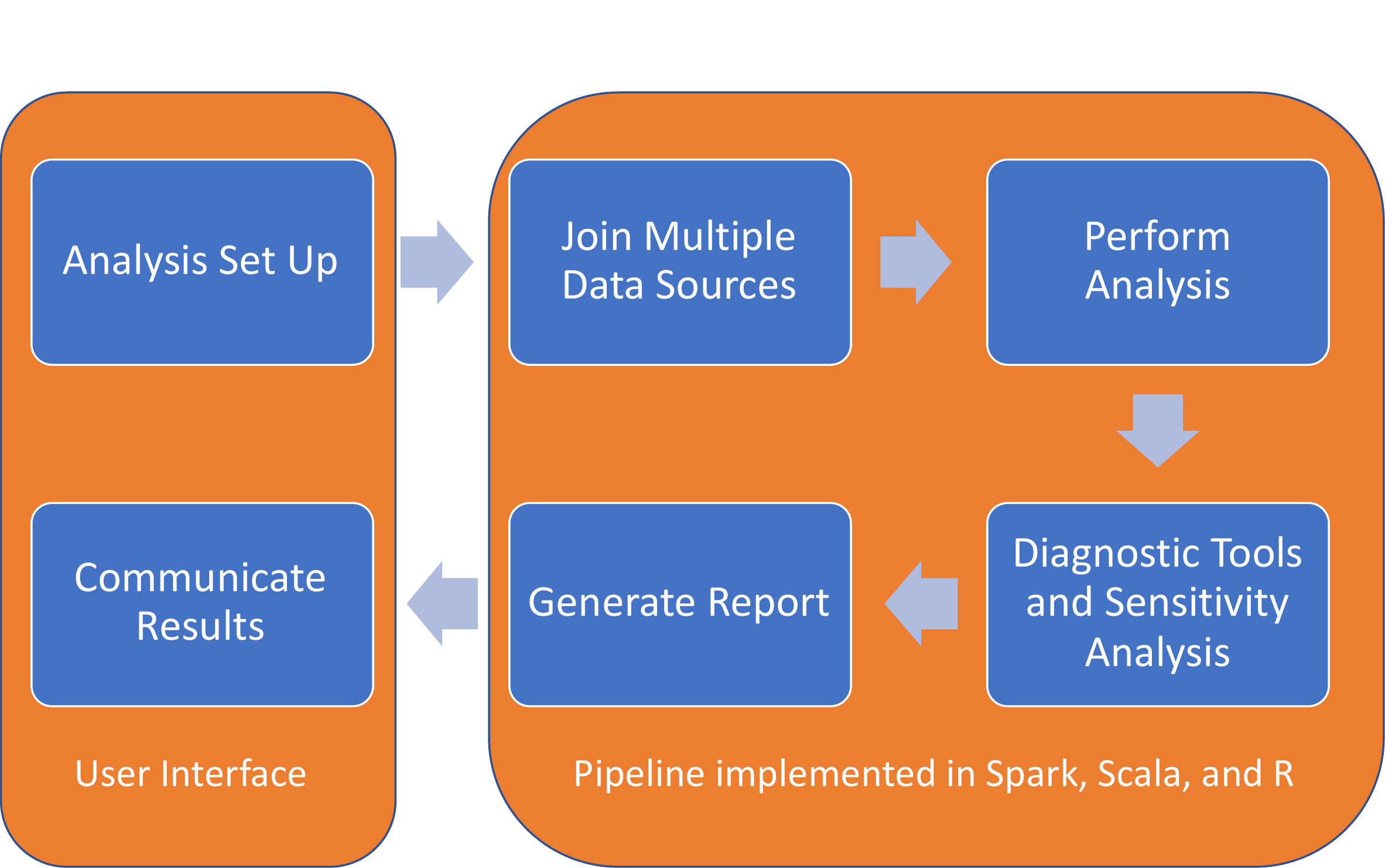}
    \caption{Flow chart of the LinkedIn causal inference platform. The modular design allows for simple onboarding of new causal methods.}
    \label{fig:flow}
\end{figure}

To allow for the democratization of causal inference at LinkedIn, we developed an internal observational study platform and integrated it with our existing experimentation platform. Our guiding principle during the development was to build a tool that made the analysis transparent, standardized, secure and accessible to data scientists. To keep the analysis transparent we store every iteration preventing that users from optimizing their results. We also implement multiple diagnostic checks and have a group of experts review important analysis before making impactful business decisions to ensure that the study met our high-quality standard.  By providing a centralized platform, we ensure that our data scientists only use stable methods that have been vetted by our team. At LinkedIn we take the utmost care to ensure the privacy of our members; therefore, most member data is not accessible by employees, by having a platform perform the analysis we are keeping our members' data safe as it is never directly seen by the analyst. Finally, by building an intuitive user interface and requiring minimal user input, we ensure that every data scientist can take their analysis to the next level. 
Figure \ref{fig:flow} illustrates the key stages in the platform.

\bibliography{reference} 


\end{document}